# Yellow whispering-gallery-mode lasing from amorphous fluoride microspheres

**ABHISHEK SURESHKUMAR,**[1] **JONATHAN DEMAIMAY,**[2] **GEORGES PERIN,**[1] **CHRISTELLE VELLY,**[1] **HELENE OLLIVIER,**[1] **YANNICK DUMEIGE,**[1] **ALAIN BRAUD,**[2] **PATRICE CAMY,**[2] **STÉPHANE TREBAOL,**[1,*] **AND PAVEL LOIKO**[2]

[1]*ENSSAT, UMR 6082-CNRS-Institut FOTON, Université de Rennes, 6 rue de Kerampont, 22300 Lannion, France*
[2]*Centre de Recherche sur les Ions, les Matériaux et la Photonique (CIMAP), UMR 6252 CEA-CNRS-ENSICAEN, Université de Caen Normandie, 6 Boulevard Maréchal Juin, 14050 Caen Cedex 4, France*
[*]*stephane.trebaol@univ-rennes.fr*

**Abstract:** Compact, low-noise coherent light sources in the visible remain challenging due to limited gain platforms and inefficient pumping. We report a new route to visible microlasing based on direct, one-photon blue pumping and an amorphous fluoride gain material platform. Dysprosium doped fluoride microspheres are fabricated via plasma-torch-induced, pressureless amorphization of single crystals, enabling compositions beyond conventional glass-forming limits while ensuring ultrasmooth morphology, low phonon energy, and homogeneous dopant distribution. We demonstrate the first fiber-coupled whispering-gallery-mode lasing from an amorphous fluoride microsphere in the yellow (573 nm), with an ultralow threshold of 190 µW despite spin-forbidden $Dy^{3+}$ transitions. Lasing is evidenced by characteristic light-light curve indicating a low spontaneous emission factor ($\beta = 3.45 \times 10^{-4}$), narrow-linewidth emission, and relaxation oscillations yielding a loaded quality factor of $Q = 3.5 \times 10^6$. This platform is readily extendable to other rare-earth emitters, enabling entire visible spectral coverage beyond the limitations of upconversion pumping, with prospects for color-tunable and white-light emission. Finally, fiber-based amplification of the WGM signal demonstrates a pathway toward compact, fiber-integrated visible microlasers with controllable noise and linewidth.

## 1. Introduction

The current trend in photonics is miniaturizing coherent light sources down to the microscale. Confining light to microscale cavities brings the advantages of i) extremely small footprint enabling dense integration, ii) low laser thresholds and enhanced efficiency, iii) fast modulation speeds, iv) high spectral purity and low noise and v) small mode volumes leading to enhanced light-matter interaction [1-3]. Microcavity lasers find applications in high-density optical data storage [4], high-speed signal processing [5], lab-on-a-chip diagnostics and biosensing [6,7], high-resolution spectroscopy [8], nonlinear optics [9,10], quantum photonics [11], and nano-optomechanics [12]. Micro-lasers can operate based on mechanisms such as Fabry-Perot cavities [13], whispering-gallery-mode (WGM) resonators [14-17], photonic-crystal cavities [18], and distributed-feedback structures [19]. Random lasers can also be included under the broader category of micro-lasers, especially when emphasizing size and coherent emission, even though the underlying physical mechanism (multiple scattering in disordered medium) is different.

Whispering-gallery-mode microcavities are optical resonators in which light circulates along the curved periphery via continuous total internal reflection, enabling high-quality-factor ($Q$), low-mode-volume confinement of light and high finesse. Geometrically, they are realized in the form of microspheres, microtoroids or microdisks [1]. Microspheres benefit from very high achievable $Q$-factors [20], tight mode confinement near the sphere surface enhancing the light-matter interaction, strong evanescent field and a broad operation spectral range at the expense of more challenging light coupling and integration as compared to lithographically defined cavities. Remarkably, WGM microsphere resonators feature narrow linewidths: the light is trapped for a long time in a low-loss, highly symmetric cavity resulting in ultra-high spectral selectivity [21,22].

Traditionally, WGM microresonators have been mainly exploited in the infrared, due to the availability of gain media (*e.g.*, glasses doped with rare-earth ions), low operation thresholds and accessible pump sources, as well as weak light scattering scaling as $1/\lambda^n$ retaining high $Q$-factors [23-25]. In particular, $Er^{3+}$-doped materials emitting at 1.5 µm attracted the most attention for low-threshold, narrow-linewidth microlasers for telecommunications, *i.e.*, dense wavelength-division multiplexing [26,27]. The realization of efficient micro-lasers operating in the visible spectral range remains an active area of research. Such light sources are of particular interest for full-color micro-displays, quantum optics, underwater communications, and high-resolution fluorescent imaging, where visible emission enables higher spatial resolution and compatibility with common detectors and fluorescent markers.

Until now, the primary approach to generate visible light from WGM micro-lasers has relied on frequency upconversion under infrared pumping, via multiphoton processes such as excited-state absorption, energy-transfer upconversion, or photon avalanche [28,29]. In this regard, rare-earth ions known as infrared emitters, notably $Er^{3+}$, $Tm^{3+}$ and $Ho^{3+}$, were employed, cf. Fig. 1(a,b). The feasibility of upconversion pumping resulting in anti-Stokes emission of the considered rare-earth activator ions is enabled by their ladder-like energy-level schemes with multiple resonant transitions. As a state-of-the-art result, the work of von Klitzing *et al.* demonstrated a green upconversion micro-laser based on an $Er^{3+}$-doped fluoride glass microsphere, achieving a $Q$-factor of $10^6$ and a remarkably low lasing threshold of 30 µW [30]. Recently, Jiang *et al.* significantly advanced this approach by demonstrating simultaneous lasing in the ultraviolet, visible, and near-infrared spectral regions using $Er^{3+}$,$Yb^{3+}$-codoped silica microspheres [31]. Owing to their ultra-high $Q$-factor, these devices exhibited lasing thresholds down to 1 µW.

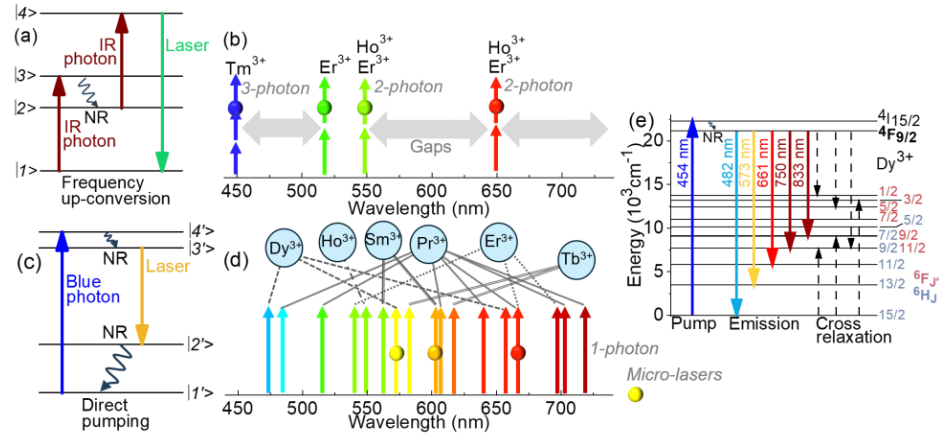

**Fig. 1**. Routes for generating visible light in $RE^{3+}$-ion-based micro-lasers: (a,b) *Upconversion pumping*: (a) excitation mechanism and (b) reported visible laser lines arising from *n*-photon processes; (c-d) *Direct pumping*: (c) excitation mechanism, (d) a nearly gapless coverage of the visible spectral range achievable through electronic transitions of $RE^{3+}$ ions in fluoride hosts (example: $LiYF_4$); (e) energy-level diagram of $Dy^{3+}$ illustrating excitation pathways, visible emissions, and cross-relaxation processes.

Intrinsically, upconversion-pumped visible light sources suffer from limited efficiency, high lasing thresholds, multiple competing relaxation pathways arising from additional energy resonances, and limited spectral coverage [32]. Moreover, their operation generates a higher heat load, which complicates integration into densely packed photonic circuits. Direct one-photon pumping leading to Stokes emission therefore represents an attractive alternative, cf. Fig. 1(c). Recent breakthroughs in visible laser development based on selected rare-earth ions featuring higher-lying metastable manifolds and multiple emission pathways across the visible spectrum (such as $Pr^{3+}$, $Tb^{3+}$, $Dy^{3+}$, and $Sm^{3+}$) have been largely driven by the advent of efficient blue GaN-based diode lasers [33]. Figure 1(d) provides an overview of the accessible laser lines, using the $LiYF_4$ fluoride host crystal as an illustrative example. However, to date, very little attention has been paid to the direct pumping of visible WGM microresonators. Hayakawa *et al*. reported on a $Sm^{3+}$-doped borosilicate glass microsphere emitting in the orange and featuring a moderate *Q*-factor of $10^3$ under pumping by an $Ar^+$ ion laser at 488 nm [34].

The emphasis on fluoride laser hosts for visible emission is not incidental: their low phonon energies suppress multiphonon relaxation leading to high luminescence quantum yields, they exhibit high rare-earth ion solubility (which is of particular relevance for amorphous phases), a low probability of interconfigurational excited-state absorption ($4f \rightarrow 5d$), and very weak photodarkening. This combination of properties makes fluorides highly attractive materials for blue-diode-pumped visible lasers [35]. With respect to microsphere fabrication, fluoride glasses such as ZBLAN appear as suitable candidates. However, regarding fabrication workflow, they suffer from low glass transition temperatures and a strong tendency for devitrification [36]. It is thus interesting to look for other material classes in the fluoride family with advanced vibronic and spectroscopic properties.

In the present work, we particularly focus on dysprosium ions ($Dy^{3+}$). $Dy^{3+}$ (electronic configuration: $[Xe]4f^9$) exhibits a metastable state $^4F_{9/2}$ separated by a large energy gap ($\sim 7400$ cm$^{-1}$) from a group of lower-lying multiplets $^6H_J$ and $^6F_{J'}$. The transitions to these levels give rise to multicolor visible emissions, with the most intense one, $^4F_{9/2} \rightarrow ^6H_{13/2}$, falling in the yellow spectral range, see Fig. 1(e). The associated four-level laser scheme facilitates efficient population inversion and enables low lasing thresholds [37]. Moreover, its strong yellow transition directly addresses the long-standing "green gap" problem in visible light sources.

In this work, we report a fundamentally new approach to visible whispering-gallery-mode microlasing, combining direct blue optical pumping using GaN diode lasers enabling efficient Stokes emission in the visible, and an original gain platform based on dysprosium-doped amorphous fluoride microspheres. These microspheres are fabricated from single crystals via a plasma-torch-induced, pressureless amorphization process, which simultaneously yields ultrasmooth surfaces, high optical quality, and compositional homogeneity regarding both the host matrix and rare-earth distribution, together with broadband absorption and emission spectra. Under blue excitation, the microspheres exhibit intense yellow luminescence, and we demonstrate WGM lasing at 573 nm with an ultralow threshold of only 190 µW. To the best of our knowledge, this constitutes the first realization of a directly pumped, fiber-coupled visible WGM microlaser. Beyond this proof of concept, we further present a first investigation of WGM signal amplification using active optical fibers, outlining a viable and scalable route toward power enhancement and the practical deployment of visible microlaser sources.

## 2. Fabrication of fluoride microspheres from single crystals

*Crystal growth.* A $Dy^{3+}$-doped $LiYF_4$ single-crystal (1 at.% $Dy^{3+}$ in the melt) was grown by the Czochralski method using high-purity LiF (4N, *Sigma-Aldrich*) and $YF_3$ (obtained in-house by fluorination of $Y_2O_3$ (4N) using an excess of $NH_4HF_2$ at 300°C). The charge composition was adjusted to a molar ratio of 52% LiF and 48% $YF_3$ in accordance with the $LiF-YF_3$ binary phase diagram [38]. The crystal growth was performed in a vitreous carbon crucible under an $Ar^+$ $CF_4$ atmosphere at a temperature of $\sim 810$°C. The pulling rate was 0.5-1 mm/h, and the rotation speed was 2-6 rpm. The total growth duration was about one week. The crystal was then slowly withdrawn from the melt and cooled to room temperature at a rate of 30°C/h. The as-grown crystal boule (Ø3× 40mm) was transparent. Dy:LiYF$_4$ crystallizes in the tetragonal class (sp. gr. $I4_1/a$ - $C^6_{4h}$, No. 88, scheelite-type structure).

*Microsphere fabrication.* The main steps of the fabrication process are summarized in Fig. 2. The single crystal was mechanically crushed into powder using a mortar and subsequently sieved to obtain particles in the 50-100 µm size range. The sieved powder was then injected into oxygen-argon plasma generated by a radio-frequency (RF) plasma torch, where the particles melted and were reshaped into spherical droplets by surface tension. Upon exiting the plasma, the molten droplets were rapidly cooled and solidified at room temperature on a Petri dish placed beneath the torch, resulting in solid microspheres. Note that the diameter of the microspheres depends on the grain size of the initial particles which can be selected by using different sieves. Optimal plasma parameters were determined empirically to achieve high sphericity and surface homogeneity, as assessed by optical microscopy - both of which are critical for obtaining high-$Q$ whispering-gallery-mode resonances. Microspheres with the diameters between 40 to 90 µm were chosen for further experiments.

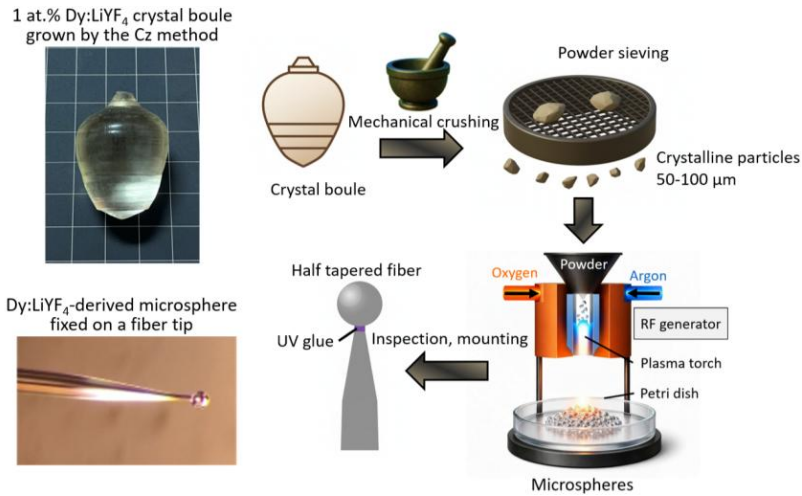

**Fig. 2**. Schematic illustrating the successive steps involved in the fabrication of fluoride microspheres using the plasma torch method. The *insets* present a photograph of the 1 at.% Dy:LiYF$_4$ crystal boule and an image of a microsphere transferred to and secured on an optical fiber half-taper.

*Mounting on fibers.* After collection and preliminary inspection, microspheres exhibiting near-perfect spherical geometry were mounted onto half-tapered silica optical fibers serving as support. A vacuum pump was used to carefully pick up an individual microsphere, which was then fixed to the fiber using a UV-curable adhesive.

### 3. Morphology and spectroscopy of Dy$^{3+}$-doped amorphous fluoride microsphere

*Methodology*. The microsphere morphology was studied using a confocal laser microscope (S-Neox, *Sensofar*) using a ×100 objective (CF60-2, *Nikon*) and a 405-nm GaN diode laser in the reflection bright-field mode. The surface topography was studied in the interference mode. µ-Raman and µ-luminescence spectra were measured using a *Renishaw* InVia Qontor confocal laser microscope equipped with an Ar$^+$ ion laser ($\lambda$ = 457, 514 nm), a ×50 *Leica* microscope objective, an edge filter and a high-resolution 2400 gr/mm diffraction grating. The spectral resolution was ~1 cm$^{-1}$. The 2D cartography was performed with a lateral resolution of 0.5 µm. A spectrofluorometer (PTI QuantaMaster-8075-22, *Horiba*) equipped with a 75 W continuous-wave lamp and a pulse Xe lamp serving as excitation sources and a photomultiplier tube

(R13456, *Hamamatsu*) was used to study the excitation spectrum and luminescence dynamics of $Dy^{3+}$ ions. The spectral resolution was 0.5 nm.

*Morphology*. Figure 3(a) presents an example confocal laser microscope image of one of the microspheres derived from the single crystal, setting the focus to the equatorial plane of the sphere. It reveals a high homogeneity of material inside the microsphere, the lack of crystallization signs, defects and bubbles. The surface presents a neatly spherical shape, revealed in the interferometric mode, Fig. 3(b).

*Evidence of material amorphization*. Raman spectroscopy is a highly sensitive probe of structural changes in solids. Figure 3(c) shows the polarized Raman spectra of the tetragonal scheelite-type, optically uniaxial $Dy:LiYF_4$ crystal, recorded in the *a*(*ij*)*a* geometry (*i, j* = $\pi$ or $\sigma$ polarizations, Porto's notation [39]). The spectra exhibit pronounced anisotropy of vibronic properties, with the most intense Raman mode at 262 cm$^{-1}$ ($A_g$ symmetry, linewidth: 7.8 cm$^{-1}$) and a maximum phonon energy of 443 cm$^{-1}$ corresponding to the $E_g$ mode. In contrast, the µ-Raman spectra of the microsphere display little dependence on either the surface position or the light polarization. They are characterized by two broad, glass-like bands: a dominant band centered at ~384 cm$^{-1}$ (linewidth: 178 cm$^{-1}$) and a higher-frequency asymmetric band at 562/594 cm$^{-1}$. This spectral signature is consistent with previous reports of irreversible pressure-induced amorphization of $LiYF_4$ at a pressure of $P$ = 40 GPa [40]. Notably, the Raman spectrum of the amorphous fluoride microsphere differs from that of state-of-the-art ZBLAN glass (maximum phonon energy ~580 cm$^{-1}$), suggesting the formation of a distinct, non-equilibrium fluoride glass phase with an even lower maximum phonon energy. The phonon spectrum of this phase rules out any incorporation of oxygen impurities which may result in the formation of oxyfluoride glassy or crystalline phases. This analysis provides compelling evidence that the plasma torch technique enables efficient pressureless amorphization, offering a powerful route for fabricating amorphous microspheres potentially applicable to a wide range of crystalline materials (oxides and fluorides).

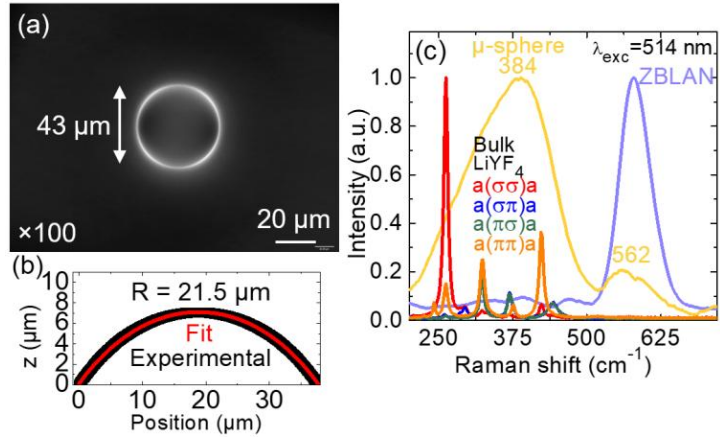

**Fig. 3.** Amorphous fluoride glass microsphere: (a) a confocal laser microscope image focusing on the microsphere equatorial plane; (b) surface topography, *black circles* - measured data, *red curve* - their parabolic fit yielding a radius $R$ = 21.5 µm; (c) Raman spectroscopy: Raman spectra of i) a $Dy:LiYF_4$ single crystal (for polarized light, *a*(*ij*)*a* geometries (*i, j* = $\pi$ or $\sigma$), ii) a fluoride amorphous microsphere derived this compound, iii) a ZBLAN fluoride glass, $\lambda_{exc}$ = 457 nm.

*Optical spectroscopy*. The photoluminescence excitation spectrum of $Dy^{3+}$ ions in the fluoride microsphere, measured in the visible spectral range, is shown in Fig. 4(a). The yellow emission at $\lambda_{lum}$ = 574 nm was monitored. The spectrum exhibits a series of broad, glass-like absorption bands corresponding to spin-forbidden ($\Delta S \neq 0$) transitions of $Dy^{3+}$ ions from the ground state $^6H_{15/2}$ to higher-lying excited manifolds. The highest excitation efficiency is expected near 390 nm, associated with the $^6H_{15/2} \rightarrow {}^4F_{7/2} + {}^4I_{13/2}$ transition. For the laser experiments, excitation at $\lambda_P$ = 454 nm was selected, dictated by the availability of the pump source and by the reduced quantum defect. This wavelength corresponds to the $^6H_{15/2} \rightarrow {}^4I_{15/2}$ transition. The associated absorption bandwidth, as large as ~10 nm, relaxes the requirement for precise wavelength stabilization, which is typically critical in crystalline $Dy^{3+}$-doped gain media.

The µ-luminescence spectrum of $Dy^{3+}$ ions in the fluoride microsphere under excitation at $\lambda_{exc}$ = 457 nm (to the $^4I_{15/2}$ state) is shown in Fig. 4(b). It depicts five emission bands in the visible, linked to transitions from the $^4F_{9/2}$ metastable state to the lower-lying manifolds $^6H_{15/2}$ (482 nm, cyan), $^6H_{13/2}$ (574 nm, yellow), $^6H_{11/2}$ (661 nm, red), $^6H_{9/2} + {}^6F_{11/2}$ (750 nm, deep-red), and $^6F_{9/2} + {}^6H_{7/2}$ (833 nm, deep-red). The yellow emission is the most intense and exhibits an emission bandwidth of 12.5 nm, Fig. 4(c). The same figure also presents the polarized emission spectra of the parent compound, Dy:LiYF$_4$ single crystal, together with those of Dy:ZBLAN glass. A direct comparison of these spectra further supports the amorphous nature of the microsphere. The luminescence decay curve from the $^4F_{9/2}$ manifold is displayed in Fig. 4(d). It evidences a certain degree of luminescence self-quenching via cross-relaxation between the neighboring $Dy^{3+}$ ions, resulting in the effective luminescence lifetime, $<\tau_{lum}>$ = $\int t I_{lum}(t)dt/\int I_{lum}(t)dt$ of 0.75 ms. This value is slightly shorter than that for the parent 1 at.% $Dy^{3+}$-doped single crystal, $<\tau_{lum}>$ = 0.93 ms, which could be assigned to a slightly higher local concentration of $Dy^{3+}$ ions due to the volatility of LiF which may change the material stoichiometry during amorphization in the plasma torch, as well as quenching by defects. The "slow" decay component is close for both materials, $\tau_{lum}$ ~ 0.94 ms. For the low-doped (0.2 mol%) Dy:ZBLAN glass, the luminescence decay is nearly single exponential yielding a lifetime of 1.29 ms.

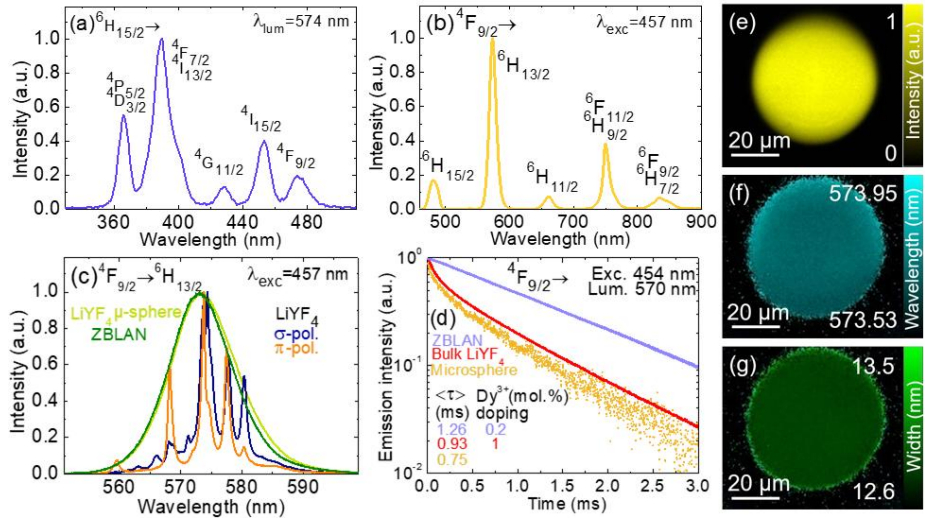

**Fig. 4.** Spectroscopy of $Dy^{3+}$ ions in fluoride microspheres: (a) excitation spectrum in the UV-blue spectral ranges, $\lambda_{lum}$ = 574 nm; (b) µ-luminescence spectrum recorded from the center of the microsphere; $\lambda_{exc}$ = 457 nm; (c) Comparison of the yellow emission band for the microsphere, the parent Dy:LiYF$_4$ crystal (π- and σ-polarized light), and Dy:ZBLAN glass; (d) luminescence decay curves from the $^4F_{9/2}$ $Dy^{3+}$ manifold for the above-mentioned materials; $\lambda_{exc}$ = 454 nm, $\lambda_{lum}$ = 570 nm; (e-g) µ-luminescence mapping across the equatorial plane of the microsphere monitoring its yellow emission at 574 nm: (e) emission peak intensity; (f) peak position; and (g) peak width.

The microsphere homogeneity was further assessed by µ-luminescence mapping, see Fig. 4(e-g). The yellow emission at ~574 nm was monitored and the band intensity, emission peak position and emission bandwidth were plotted across the equatorial plane of the microsphere. These maps confirm the highly homogeneous distribution of $Dy^{3+}$ ions and the regularity of their emission properties across the microsphere, and they rule out any signs of crystallization. The slight evolution of the emission properties near the surface (*i.e.*, peak broadening) can be assigned to residual stress.

Amorphization is the process by which a material loses its long-range, ordered crystal structure and becomes amorphous. It can be observed via rapid cooling (quenching), mechanical deformation, high pressure applied, or irradiation. Based on the combined results of the performed spectroscopic studies, we confirm the irreversible amorphization of tetragonal scheelite-type Dy:LiYF$_4$ crystals, leading to the formation of a low-phonon-energy fluoride amorphous glass-like phase that exhibits strong inhomogeneous broadening of both excitation and emission spectral profiles. This phase is stable in ambient conditions in air and does not show any degradation of its emission properties over time highlighting its potential for WGM micro-lasers.

### 4. Whispering gallery mode microlaser and amplifier seed

*Laser setup*. As a pump source, we used a blue semiconductor laser diode (*Oxxius*) emitting up to 70 mW at a central wavelength of 454 nm (linewidth: 1.2 nm) matching the $^6H_{15/2} \rightarrow \,^4I_{15/2}$ transition in absorption of Dy$^{3+}$ ions. The output was coupled into a single-mode silica fiber (S405-XP, cut-off wavelength: 380 nm, mode field diameter: 3.3±0.5 µm), and a fiber coupler split the pump power, directing 90% into a full fiber taper for coupling to the microsphere, while the remaining 10% was used for power monitoring. The fiber taper was fabricated using the SMF-28 optical fiber (core diameter: 8.2 µm) using the heat pull method resulting in the waist diameter of ~1 µm. An in-line polarization controller was incorporated into the tapered fiber to optimize the coupling of selected modes into the microsphere cavity. The fiber taper was designed not only to efficiently deliver the pump light to the microsphere, but also to collect the emitted radiation via evanescent-wave coupling. The microsphere was brought into proximity with the fiber taper until it was attracted by electrostatic forces, corresponding to the overcoupling regime, which provided high mechanical stability for long-term operation. The detection was enabled by an optical spectrum analyzer (*Yokogawa* AQ6373E, spectral resolution: 0.1 nm) and a high sensitivity fast PIN photodetector (*Menlo* FPD 510 FV, rise time: < 2 ns) coupled to a radio frequency (RF) spectrum analyzer (*Rohde & Schwarz* FSW-8GHz). A long-pass filter (LP550) was used to filter out the residual blue pump.

For the proof-of-principle amplification experiment using the yellow WGM microlaser as seed source, we connected the fiber taper to a 1-m long Dy$^{3+}$-doped silica glass fiber (*Exail*, core diameter: 25 µm, N.A. = 0.18, 0.3 mol% Dy doping). This configuration offers the advantage of simplicity: the taper output simultaneously carries both the seed laser signal and the residual blue light, which further pumps the Dy:silica fiber, enabling straightforward observation of amplification.

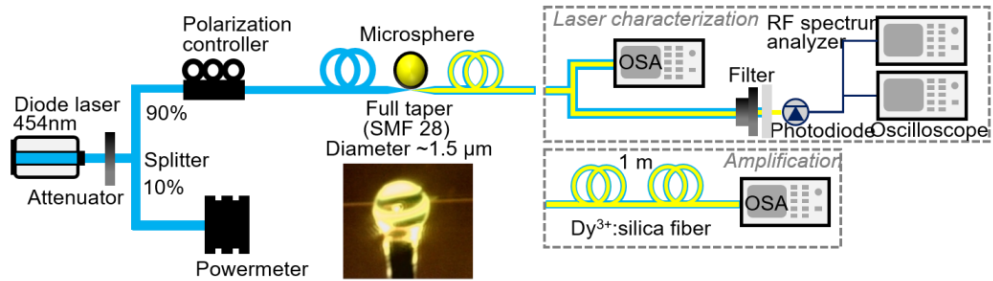

**Fig. 5.** Outline of the optical setup for laser characterization of the Dy$^{3+}$-doped fluoride microsphere, as well as proof-of-principle amplification experiment using the whispering gallery mode laser signal as a seed in a Dy$^{3+}$-doped silica fiber. *Inset* - a photograph of an optically pumped microsphere.

*Whispering gallery mode microlaser.* To prove laser action in the microsphere (and rule out, *e.g.*, amplified spontaneous emission or spectral filtering), we have involved three distinct approaches: i) observation of a threshold laser behavior, ii) observation of emission linewidth narrowing, and iii) a study of the emission dynamics, notably, relaxation oscillations [41]. Figure 6(a) schematically depicts the principle of operation of the microlaser. To ensure single-mode operation of the microlaser, the pump source temperature and driving current, as well as the coupling geometry, were first optimized. Once the optimal parameters were identified, they were fixed, and the pump power was attenuated to obtain the microlaser power transfer curve.

The power transfer characteristics of the microlaser are presented in Fig. 6(b), where the emitted yellow power collected at the exit of the full fiber taper is plotted as a function of the absorbed blue pump power, $P_{abs}$. The latter was derived from the blue pump exiting the full taper without and with the coupled microsphere. Note that the taper itself induced a power loss for the blue pump of ~5 dB due to germanium absorption in the SMF-28 fiber. A clear threshold behaviour was observed, yielding a threshold $P_{th}$ of 190 µW. The output dependence was almost linear, resulting in a maximum output of 300 nW and a slope efficiency $\eta$ of 3%. Well above threshold, a rollover in the output power is observed, which may originate from the gain saturation due to population accumulation in the terminal laser level (see below).

The evolution of the emission spectra of the $Dy^{3+}$-doped fluoride microsphere as a function of absorbed pump power, spanning operation from below to above lasing threshold is presented in Fig. 6(c-d). Below threshold, the spectra are dominated by broadband amplified spontaneous emission (ASE). As the absorbed pump power increases toward threshold, the emission evolves into a dominant narrow peak, indicating the onset of laser oscillation (note that it cannot be fully resolved with the used equipment limiting the spectra resolution to 0.1 nm). When plotted in a logarithmic scale, the spectra reveal an important increase of the signal-to-noise ratio (reaching 27 dB above the noise level for $P_{abs}$ = 0.30 mW) as well as a modulation of the ASE background by the resonances of whispering-gallery-modes.

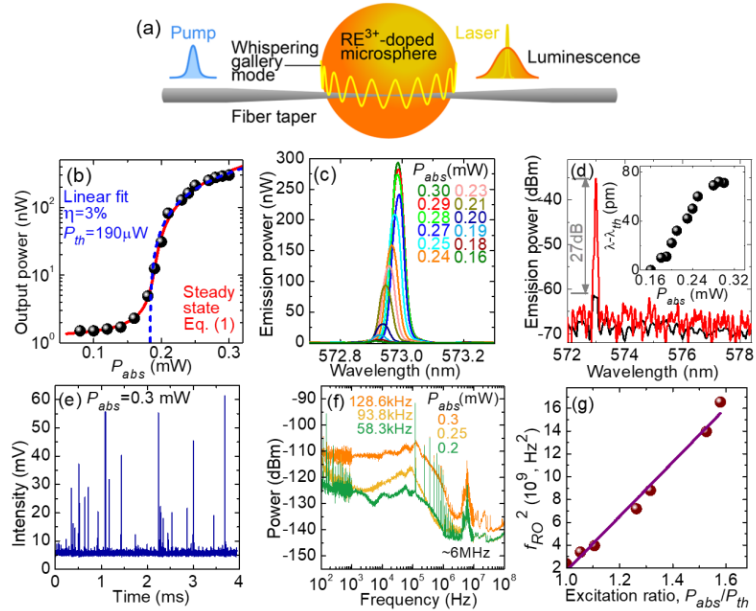

**Fig. 6.** Yellow whispering gallery mode laser based on a $Dy^{3+}$-doped fluoride microsphere: (a) A scheme representing energy conversion of the blue pump light coupled into the microsphere via a fiber taper, leading to WGM laser emission; (b) power transfer curve: *symbols* - experimental data, *curve* - their fit using Eq. (1), $P_{th}$ - laser threshold, $\eta$ - slope efficiency; (c,d) spectra of laser emission measured at different pump levels: (c) linear scale; (d) log scale: *black curve* - below threshold, $P_{abs}$ = 0.18 mW, *red curve* - well above the threshold, $P_{abs}$ = 0.30 mW,

*inset* - power-dependent laser emission peak shift; (e) a typical oscilloscope of laser emission; (f) intensity-noise measurements for varying pump level; (g) squared frequency of relaxation oscillations plotted versus the excitation ratio, *circles* - measured data, *curve* - their linear fit.

In addition to spectral narrowing, a systematic redshift of the emission peak is observed with increasing pump power. The laser line undergoes a monotonic redshift of approximately 80 pm across the investigated excitation range, see the inset of Fig. 6(d). This shift is attributed to thermally induced effects in the microsphere, including temperature-dependent changes in the refractive index ($dn/dT$ and stress-optic terms) as well as volumetric thermal expansion of the microresonator [42].

The yellow output from the microlaser exhibited characteristic spiking behavior, see the example oscilloscope trace in Fig. 6(e). Self-pulsing is frequently observed in visible lasers based on rare-earth ions [43] and it can be explained by "fast" saturable absorption related to excited-state absorption from short-living intermediate energy levels. In the case of Dy lasers operating on the $^4F_{9/2} \rightarrow {}^6H_{13/2}$ transition, the population is accumulated in the terminal laser level ($^6H_{13/2}$) which has a relatively long lifetime, especially in low-phonon energy fluoride compounds ($\tau_{lum}$ ~400 μs for Dy:LiYF$_4$). It can act as a reservoir for populating other intermediate states via different cross-relaxation mechanisms. Self-pulsing can be suppressed by codoping the laser material with Tb$^{3+}$ ions efficiently quenching the $^6H_{13/2}$ manifold via non-radiative energy transfer and multiphonon relaxation [44].

We further performed intensity noise measurements for the designed microlaser, Fig. 6(f). For an absorbed pump power of 300 μW, the spectrum shows a distinct low-frequency peak at 128 kHz assigned to relaxation oscillations. A broad and intense high-frequency peak at 6 MHz could be attributed to beating between the clockwise and counterclockwise lasing modes circulating in the cavity. The beating peak provides compelling evidence of lasing, as it appears only under active lasing conditions [47]. Additional weaker peaks were attributed to other cavity modes.

Due to the relatively long upper laser level lifetime of Dy$^{3+}$ ions, our WGM microsphere laser can be regarded as a class B laser: $\tau_\perp \ll \tau_c \ll \tau_i$, where $\tau_\perp$ is the polarization (dipole) relaxation time, $\tau_c$ is the photon lifetime in the cavity, and $\tau_i$ is the population inversion lifetime, meaning that the photon density and population inversion are both dynamic variables and relaxation oscillations are present [45]. The noise characteristics of the microlaser (the frequency of relaxation oscillations $f_{RO}$) were recorded at various pump powers, as shown in Fig. 6(g). One can see that $f_{RO}^2$ depends almost linearly on the ratio between the absorbed pump power and the microlaser threshold power (the excitation ratio, $r = P_{abs}/P_{th}$). This dependence was then fitted using the formula $f_{RO}^2 = \frac{r-1}{\tau_c \tau_i}$. We assume that $\tau_i \approx \tau_{lum}$. This is valid because in the case of Dy$^{3+}$ in the fluoride matrix, the radiative decay from the upper laser level dominates, and due to the relatively low doping level, the cross-relaxation phenomena are still weak. By fitting the experimental data, the cavity lifetime was determined to be $\tau_c$ = 1.08±0.07 ns, corresponding to a loaded quality factor of $Q = (2\pi c/\lambda_0)\tau_c$ = (3.5 ± 0.23)×10$^6$ ($\lambda_0$ - resonance wavelength). In this wavelength range, the quality factor is generally limited by surface roughness, which makes the cavity more sensitive to Rayleigh scattering at short wavelengths. For example, for a silica microsphere, a roughness-limited quality factor close to 10$^8$ is expected [46]. Nevertheless, laser emission in WGM microspheres often implies the excitation of high-order modes [47], described by a large mode volume. Indeed, the overlap of the lasing mode with surface irregularities is increased and may explain the relatively modest quality factor observed experimentally. Moreover, as mentioned earlier, the microsphere is operating in the over-coupling regime which could also explain the moderate loaded Q-factor. The cold-cavity linewidth of the microsphere resonance can be estimated from its quality factor as $\Delta\nu = \nu_0/Q$ ~ 150 MHz. Above threshold, the laser linewidth is expected to be much narrower due to stimulated emission.

*Modeling of spontaneous emission coupling factor.* The spontaneous emission factor ($\beta$) is the fraction of all spontaneous emission that is coupled into the lasing whispering gallery mode.

This parameter can be estimated by analyzing the pump-output characteristic using a semi-classical rate-equation model [48,49]. The cavity photon number *s* is governed by the steady-state equation:

$$s = -\frac{1}{2\beta} + \frac{p}{2\Omega} + \frac{\sqrt{(\beta p - \Omega)^2 + 4p\beta^2\Omega}}{2\Omega\beta}, \quad (1)$$

where $p$ is the normalized pump rate and $\Omega = 2\pi/\tau_c$ denotes the cavity loss rate, independently determined from the relaxation oscillation frequency of the cavity. The solid line in Fig. 6(b) corresponds to a three-parameter fit ($s$, $p$, $\beta$) of the experimental data based on Eq. (1). An additional offset parameter is included to account for the finite noise floor of the detection system, evaluated at 1.26 nW. The fitting procedure converges to a unique parameter set, yielding an estimated spontaneous emission factor of $\beta = 3.45 \times 10^{-4}$. Such a low value of $\beta$ can be attributed to the broad emission bandwidth of $Dy^{3+}$ ions (12.5 nm, see Fig. 4(c)), which is distributed over the numerous WGM families supported by the microsphere resonator geometry. This small $\beta$ factor is favorable for the realization of ultra-low-noise microlasers. However, direct noise characterization remains challenging due to the low output power of the microlaser, which is below the sensitivity threshold of current frequency noise measurement systems.

*Fiber amplifier seeded by a microlaser.* One notable limitation of whispering-gallery-mode microlasers is their intrinsically low output power, particularly in the visible. Nevertheless, their exceptionally narrow emission linewidths and compatibility with fiber-based architectures make them ideal candidates as seed sources for fiber master-oscillator power amplifiers (MOPA). For a proof-of-principle demonstration, we employed the experimental configuration depicted in Fig. 5. $Dy^{3+}$ ions in the silica fiber present a broad emission in the yellow centered at 578 nm with a bandwidth of ~23 nm, well matching the spectrum of the seed microlaser. We used the full available power from the 454 nm blue pump source (70 mW). At the output of the full fiber taper, the residual blue pump power was measured to be 41 mW, which was then coupled into the $Dy^{3+}$-doped silica glass fiber, serving as the amplifying medium. The pump absorption in the latter was nearly complete. The yellow microlaser emission (~300 nW) was also coupled into the active fiber.

By comparing the spectral power density of the yellow emission measured with the OSA at the output of the full fiber taper and at the output of the Dy:silica fiber, a small-signal gain of 19 dB was obtained, with a WGM laser input power of -43.9 dBm. Despite the measured noise figure of 5.16 dB, the amplifier maintains acceptable signal quality, confirming its suitability for seeding and further amplification stages. The proposed concept of a whispering-gallery-mode seed microlaser combined with a fiber amplifier [50] could be further developed by optimizing the active fiber material (*e.g.*, silicate or fluoride - Dy:ZBLAN [37,51]) and its geometry to minimize coupling losses and maximize gain. Note that ZBLAN glass fibers are much less prone to photodarkening under blue light irradiation as compared to their silica counterparts.

## 5. Conclusions

In conclusion, we introduce a novel strategy for the realization of low-threshold, narrow-linewidth coherent light sources in the visible spectral range. This approach leverages the synergy between direct blue pumping using GaN diode lasers via a Stokes, one-photon excitation process, and an original amorphous fluoride gain platform enabled by the pressureless amorphization of rare-earth-doped fluoride single crystals using a plasma-torch technique. On the materials side, this method unlocks compositions that are otherwise inaccessible with conventional glass technology, overcoming the intrinsically narrow glass-forming domain of fluorides. The resulting microspheres combine spherical morphology, low surface roughness, low phonon energy, homogeneous rare-earth ion distribution, and spatially uniform optical properties, confirmed by µ-spectroscopy, together with broadband absorption and emission characteristics.

These advances culminate in the first demonstration of whispering-gallery-mode lasing from an amorphous fluoride microsphere in the yellow spectral region, based on dysprosium ions, an emission that remains particularly challenging to achieve in conventional bulk laser geometries. The WGM architecture, characterized by intrinsically strong light-matter interaction within the cavity, enables an ultralow lasing threshold of 190 µW despite the spin-forbidden nature of $Dy^{3+}$ optical transitions, resulting in weak absorption. Clear evidence of lasing is provided by the light-light power curve, accurately described by a model incorporating the spontaneous emission coupling factor, as well as by narrow emission spectra with a side-mode suppression ratio (27 dB above the noise floor). In addition, the observation of relaxation oscillations allows us to estimate a loaded quality factor $Q$ of $(3.5 \pm 0.23) \times 10^6$.

The proposed approach is readily extendable to other rare-earth ions supporting direct visible emission in fluoride hosts, opening a pathway toward near-complete spectral coverage across the visible, unlike upconversion schemes, which are limited to a few discrete emission bands. Furthermore, ions exhibiting multiple transitions with comparable gain (such as $Pr^{3+}$) could enable color-tunable operation, potentially extending to white-light emission.

Finally, we demonstrate a proof of principle for the amplification of the narrow-linewidth WGM laser signal using a rare-earth-doped optical fiber, pointing toward a new paradigm of fiber-integrated, compact visible light sources with controllable noise and spectral linewidth. Further control over the excitation geometry and modal behavior within the microresonator is expected to enable polarization engineering of the WGM emission, providing an additional and powerful degree of freedom in laser design.

**Funding.** Indo-French Centre for the Promotion of Advanced Research (70T12-2, MASSALAQ), I-DEMO PIA4 Bpifrance - Région Bretagne - Lannion Tregor Communauté (40898814/1, QoQeliQo), Lannion Tregor Communauté – Région Bretagne (ARED) (ELVIS Project), Région Normandie, France (Contrat de plan État-Région); Agence Nationale de la Recherche (ANR), France (ANR-22-CE08-0025-01, NOVELA).

**Acknowledgment.** The authors thank *Oxxius* for providing the laser diode and *Exail* for supplying the Dysprosium-doped fiber used for the optical amplification.

**Disclosures.** The authors declare no conflicts of interest.

**Data availability.** Data underlying the results presented in this paper are not publicly available at this time but may be obtained from the authors upon reasonable request.